\documentclass[prd,onecolumn,groupedaddress]{revtex4-1}

\usepackage{amsmath}
\usepackage{amsfonts}
\usepackage{amssymb}
\usepackage{epstopdf}
\usepackage{amsthm}

\begin{document}

\title{On the relation between the monotone Riemannian metrics on the space of Gibbs thermal states and the linear response theory}
\author{ N. S. Tonchev}

\affiliation{
 Institute of Solid State Physics, Bulgarian Academy of Sciences,
1784 Sofia, Bulgaria}

\begin{abstract}

The proposed  in J.  Math. Phys. v.57, 071903 (2016)  analytical expansion of monotone (contractive) Riemannian metrics (called also quantum Fisher information(s)) in terms of moments of the dynamical structure factor (DSF) relative to an original intensive observable is reconsidered and extended.  The new approach through the DSF which characterizes fully the set of monotone Riemannian metrics on the space of Gibbs thermal states is utilized to obtain an extension of the spectral presentation obtained for the Bogoliubov--Kubo--Mori metric (the generalized isothermal susceptibility) on the entire class of monotone Riemannian metrics. The obtained spectral presentation is the main point of our  consideration. The last allows to present the one to one correspondence between monotone  Riemannian metrics and operator monotone functions (which is  a  statement  of the Petz theorem in the quantum information theory) in terms of  the linear response theory. We show that monotone Riemannian metrics can be determined from the analysis of the infinite chain of equations of motion of the retarded Green’s functions. Inequalities between the different metrics have been obtained as well. It is a demonstration that the analysis of information-theoretic problems has benefited from concepts of statistical mechanics and might cross-fertilize or extend both directions, and vice versa. We illustrate the presented approach on the calculation of the entire class of monotone (contractive) Riemannian metrics on the examples of some simple but instructive systems employed in various
physical problems.

\end{abstract}


\maketitle



\section{Introduction}
During the last five years  there is an increasing interest in deriving  relations between two seemingly unrelated fields, i.e. the
metric space on the set of quantum states (information geometry) and linear response of thermal systems (statistical mechanics)
\cite{GY14,YH15,T16,SU16,HHTZ16,KSMP17,F18,LVSC19,CSDV19,k19,CVS20}.

Illustrative example represents
such a geometry-based notion as  fidelity susceptibility \cite{Gu10} (and refs.therein).
It is defined as a coefficient (in front of the second term) in the expansion of a measure of  distinguishability between two points on the manifold of density matrices, and  as firstly  noted in \cite{GY14} in the ground state is potentially measurable in experiments.
To avoid confusion let us note that the fidelity susceptibility appears in various contexts under different names. Namely, it  equals to the Bures-Uhlmann metric (based on the
symmetric logarithmic derivative (SLD)) which is, however, proportional, up to a factor 1/4, to the minimal quantum Fisher  information \cite{BC94,P2009}. More precisely a detail study  of the reasons of the break down of the continuity connection between the
quantum Fisher information and the Bures metric
(or the fidelity susceptibility)  has been studied in \cite{Sa17}.
Two nonequivalent definitions of  fidelity
susceptibility have been used in the literature:
one based on the Uhlmann's  fidelity (see, e.g. \cite{Gu10}) and  another   based on  ``fidelity'' introduced in \cite{S10} and having a presentation in terms of nonzero temperature  Green's functions \cite{AASC10}, see also  the Discussion in \cite{BT12}.  Notice that  in the ground state both definitions coincides.  Equalities that relate zero-temperature fidelity susceptibility and either zero-momentum DSF \cite{GY14} or negative-two-power moment of DSF \cite{YH15} make fidelity susceptibility   
 an experimentally measured quantity. In ref.~\cite{YH15}  it was announced  that the relation between the ground state  fidelity susceptibility and the negative-two-power moment of DSF  may be extended to the finite temperature case in the spirit of ref.~\cite{AASC10}.
The major result of ref.~\cite{HHTZ16} is the established frequency integral presentation of the Fisher  information through the dissipative part of the dynamic susceptibility.
In \cite{HHTZ16} the scenario that renders density matrices $\rho_1$ and $\rho_2$  distinguishable is due to a unitary transformation generated by a hermitian operator  associated with the parameter under estimation.

Thus, in the above cited works, links have been found between two basic geometrical quantities~-- the quantum Fisher informations regarded as monotone Riemannian metrics, and the dynamical structure factor (DSF)  at both zero and finite temperature.
 The established neat relations unequivocally show that the former is not merely a theory of information topic but may have  implications on physical experiments as well.

 In informal terms, the underlying idea one follows  is to endow
 the set of quantum  states (i.e. the space of density matrices $\rho$)   with a metric structure~-- a smoothly varying positive definite inner product on the tangent spaces at a point $\rho$ and thus explored as a Riemannian manifold, see e.g. \cite{BZ06,PG11,DS14}. Due to the non-commutativity nature of the density matrices there is no unique solution of this problem. In the geometrical approach to statistics, which does distinguish between classical and quantum probabilities, proposed by Morozova and $\breve{C}$encov~\cite{MC90}, and Petz \cite{P96} the entire  class of  the monotone (or contractive) Riemannian metrics (quantum Fisher information(s)) can be introduced and studied from a unified point of view on the basis of the established  one-to-one correspondence between the monotone Riemannian metrics (MRM) and a special class of  L$\ddot{o}$vner operator monotone functions. The abundance of metrics raises the interesting question of their potentially importance in the context of quantum statistical mechanics and condensed matter physics.
The  isothermal susceptibility \cite{PT,P94,K57,M65,B61,R09} and  the quadratic fluctuations (the variance) of a quantum observable \cite{J86,P94} are prominent examples that
  have a clear interpretation in terms of Riemannian geometry on the state of space.
 Therefore, it seems quite natural a similar relationships to be looked for on the entire class of Riemannian metrics.
In fact this point of view has been adopted, albeit in a different way, in refs. \cite{R09,R13,T16,SU16,F18}.

It is interesting to make a special comment on the works \cite{T16,SU16} and \cite{F18} because of the obtained complementary
 results in a common field.

  A new presentation of the monotone Riemannian metrics (using the  Morozova, $\breve{C}$encov and Petz classification) in terms of the frequency moments of DSF  has been investigated in \cite{T16}. This approach allows to evaluate the metrics by an expansion based on the sum rules of the frequency moments of DSF. In some important cases, due to the symmetry properties of the considered model, the proposed expansions may be evaluated in a closed analytical form.

In \cite{SU16} a generalized version of the fluctuation-dissipation theorem, which relates response functions to generalized covariances (introduced earlier \cite{GHP09} and are nothing but quantum Fisher informations) has been obtained and explored. On the basis of this result a method to determine the generalized covariance from the admittance of the dynamical susceptibility has been developed.

Quite a different approach to quantum correlations which is not motivated by
 geometrical ideas has been developed in \cite{F18}. The quantum Fisher information (QFI) and the quantum variance were considered as members of
 a wider family of coherence measures. They both quantify the
 speed of evolution of the state under a unitary transformation, although for different
 measures of distinguish-ability between the states of the system. At thermal equilibrium, all the coherence
 measures of this family were expressed in terms of the dynamical susceptibility. As a result a metric approach to phase transitions has been constituted.

 It is worth noting that, in the above commented approaches the starting idea and the statistical model which renders the set of density matrices  are different.
 In  \cite{T16}  a term  added to the given Hamiltonian  which parametrized the Gibbs thermal states of the system (Gibbs statistical model) is used, accenting on the relation with statistical mechanics. In \cite{SU16,F18} the family of density matrices is obtained via  a unitary transformation generated by a Hermitian operator (unitary statistical model) in the context of the parameter estimation theory, respectively.

 In the three works \cite{T16,SU16,F18} notable relations between notions from the linear response theory of quantum statistical mechanics and informational-
 geometric approach to quantum correlations have been established in quite  different aspects. It is the aim of the present study to explore these relations in details focusing on quantum states in exponential form (Gibbs thermal states). Recall that,  important relations in the field about the interplay between Hilbert space geometry, thermodynamics and quantum estimation theory have already been studied (see, e.g refs. \cite{ZVG07, ZPV08}).

The paper is organized as follows: In Sectin~\ref{ts2} some needed notations and basic setting concerning MRM
are presented.
In Section~\ref{ts3}, we derive a spectral presentation, which allows to relate  the entire class of the MRM to linear response functions such as
the DSF and the dynamical susceptibility. Based
on this relation and using the Green's functions method, we show in Section~\ref{ts4} how to determine
the MRM  from the dissipative component  of the Kubo response  function to an external field. In Section~\ref{ts5}, we apply our method
to some particular metrics, e.g. Bogoliubov-Kubo-Mori metric, Morozova-$\breve{C}$encov metric,  Bures (or SLD) metric and the family of Wigner-Yanase-Dyson metrics. The presentations of these MRM within a thermodynamic setting  have been
studied which allows to obtain some new inequalities between them.  We
show the applicability and the efficiency of our method in generating inequalities between different MRM in Section~\ref{ts6}.
In Section~\ref{ts7}, the presentation of the MRM in terms of  the  moments of the dynamical structure factor (DSF) and the relation with the results of Section~\ref{ts4} have been discussed.
  In Section~\ref{ts8} our approach is presented and tested on two models: system of $N$ spins in a constant  magnetic field $h$ and a model Hamiltonian which
  is employed in various physical problems such as  the displaced and single-mode squeezed
  harmonic oscillators. A summary and discussion are given in Section~\ref{ts9}.  \ref{tAA}  contains a list of most popular operator monotone
  functions.

\section[]{Notations and Basic Setting }\label{ts2}

 A MRM is a family of inner products on the tangent space of a smooth manifold that are used to measure distances on the manifold.
For future references, we need to  recall briefly the
definition of the  MRM defined on
the differential manifold formed by the quantum statistical density matrices \cite{P96,PG11,DS14,BZ06}.

A density matrix $\rho$ (known in the mathematical literature as positive trace-class operator with unit
trace-norm, see \cite{Z20} for a rigorous definition) represents the state of a quantum system associated with a Hilbert space $\mathcal{H}$. For a $n$-dimensional state it is $n\times n$  non-negative trace-one Hermitian matrix. The set of all density matrices under consideration is denoted by
\begin{equation}
\mathcal{D}(\mathcal{H})=\left\{ \rho(h) \in \mathcal{M}(\mathcal{H}):\text{Tr}\rho(h)=1, h\in G\right\},
\end{equation}
where $\mathcal{M}(\mathcal{H})$ is a
 differentiable manifold structure connected  with the algebra of  $n\times n$ matrices $M_{n}$.
Formally $h \in G$, where $G \subset \mathbb{R}$ is an open set including $0$.

For the manifold  $\mathcal{M}(\mathcal{H})$ of quantum states, the tangent space  $T_{\rho}\mathcal{D}(\mathcal{H})$ at each point $\rho \in \mathcal{D}(\mathcal{H})$  can be identified with the (real) vector space $\mathcal{B}^0_{s.a}$
of self-adjoint
operators on $\mathcal{H}$ with zero trace:
$T_{\rho}\mathcal{D}(\mathcal{H}):=\{A\in M_{n}:A=A^{+},\text{Tr} A=0\}$,
where $A^{+}$ denotes the adjoint operator of $A$.

Let us $A$ and $B$ belong to the  tangent space $T_{\rho}\mathcal{D}$ at $\rho$ of the manifold $\mathcal{D}(\mathcal{H})$. The linear mapping $M_n\to M_n$ defined as:
\begin{equation}
 L_{\rho}(A)=\rho A, \quad R_{\rho}(A)=A\rho, \quad A\in M_n
\end{equation}
stands for the left and right multiplication by $\rho$. Obviously, $  L_{\rho}R_{\rho}=R_{\rho}L_{\rho}$ considered as matrices in $M_{n^2}$.

Let us define the binary
operation on $A$ and $B$
\begin{equation}
m_{f}(A,B):=A^{1/2}f(A^{-1/2}BA^{-1/2})A^{1/2}
\label{KA}
\end{equation}
known as  the Kubo--Ando operator mean \cite{KA80}, see also  Eq.~(5.36), Chapter V in ref. \cite{HP14}, where $f(x)$ is an {\it operator monotone  function} and $\langle A,B \rangle_{HS}:=\text{Tr}(A^*B)$ is the Hilbert-Schmidt inner product. Using the notion of matrix mean one may define the class of monotone metrics (called also quantum Fisher informations) parametrized by the functions $f$.

 The existence of a wide class of MRM (quantum Fisher informations) on the
quantum statistical manifold $\mathcal{D}(\mathcal{H})$ is the essence of the quantum
Petz theorem \cite{P96} (see also  \cite{PG11,DS14}).
 The theorem states that
 formula
 \begin{equation}
 g_{\rho}^f(A,B)=\langle A,m_{f}(L_{\rho},R_{\rho})^{-1}(B)\rangle_{HS},
 \label{AB}
 \end{equation}
 establishes a one to one correspondence between MRM and a special class of operator monotone functions (named also standard and denoted by $\mathcal{F}_{op}$)   $f(x):(0,+\infty) \rightarrow (0,+\infty); f(x^{-1})=f(x)/x$  and $f(1)=1$.

 Some important examples of  $f(x)\in \mathcal{F}_{op}$ are given in the Appendix  (see also refs.
  \cite{P96,HP14,F08}). Hereafter the subscript $f$ stands for a
monotone metric which depends on the standard operator monotone function $f$.

The positive definite  bilinear form, Eq.~\eqref{AB}, defines a Riemannian distance $d$, which is such that the square infinitesimal distance $ds^2=d^2({\rho,\rho+d\rho})$ between two neighboring density operators $\rho$ and $\rho+d\rho$. In a basis independent form it is given by \cite{P96} (see also \cite{DS14}):
\begin{equation}
d_f^2=g^f_{\rho}
(d\rho,d\rho)\,.
\label{sdk}
\end{equation}

Denote the spectral decomposition of $\rho$ as $\rho=\sum_{n=1}^d \rho_n|n\rangle\langle n|$, where $\rho_n$ and $|n\rangle$ are the $n$-th  eigenvalue and eigenstate of $\rho_n$, and $d$  is the dimension of the support of $\rho$. Observe that the superoperators $L_{\rho}$ and $R_{\rho}$ commute, and thus the following relation holds
\begin{equation}
m_{f}(L_{\rho},R_{\rho})^{-1}|m\rangle\langle n |=c_f(\rho_n,\rho_m)|m\rangle\langle n|,
\label{KAm}
\end{equation}
where  the symmetric function $c_f(x,y)=c_f(y,x)$,
is called the Morozova-$\breve{C}$encov function of the metric $g_{\rho}^f$ \cite{P96}.
The Morozova-$\breve{C}$encov function has the following presentation
\begin{equation}
c_f(x,y)=
\frac{1}{x f(y/x)}.
\label{CMf}
\end{equation}
and obeys the equation $ c_f(tx,ty)=t^{-1}c_f(x,y)$
for any $t \in \mathbb{R}$.

Thus, for  $\rho$ with diagonal matrix elements $\rho_m$, through Eqs.~\eqref{AB},\eqref{sdk} and \eqref{KAm} any monotone Riemannian metric on the set of quantum states (up to a proportionality constant) is presented  as (see, e.g. \cite{BZ06,DS14,Z14,PCCAS16,k19}):
\begin{equation}
d^2_{f}=\frac{1}{4}\left\{\sum_{m}\frac{d \rho_m^2}{\rho_m} + \sum_{m,n,m\neq n}c_f(\rho_m,\rho_n)|\langle m|d\rho|n\rangle|^2\right\},
\label{CMP}
\end{equation}
where the choice of the factor $1/4$ assures the relation to Fisher information and  the consistency with the existing literature \cite{BZ06}.
In the commutative case, the Bures metric reduces to the classical Fisher information, given by the first
term.
Note that,
from now  we shall study (non-singular) full-rank
  density matrices, i.e. density matrices with all positive eigenvalues. 

We consider a set of Gibbs thermal states characterized by the one-parameter family of  density matrices:
\begin{equation}
\rho(h) = [Z_N(h)]^{-1}\exp[- H(h)], \label{roh}
\end{equation}
defined on the family of $N$-particles Hamiltonians of the form
\begin{equation}
 H(h)= T - h S,
\label{ham}
\end{equation}
where the Hermitian operators $T$ and $S$ do not commute in the general case. Here, $h$ is a real (control) parameter, $Z_N(h)= {\mathrm Tr}\exp[- H(h)]$ is the corresponding partition function and for convenience the inverse temperature $\beta$ is adsorbed in $T$ and $S$.  We assume that the Hermitian operator $T$ has a complete
orthonormal set of eigenvectors $\{|m\rangle\}$
with a non-degenerate spectrum $\{T_m\}$; $T|m\rangle = T_m|m\rangle$, where $m=1,2,\dots $ In this basis the zero-field density matrix $\rho :=\rho(0)$ is diagonal:
\begin{eqnarray}
\langle m|\rho (0)|n\rangle &=&\rho_m \delta_{m,n},\quad
\rho_m:= e^{- T_m}/Z_N(0),
 \quad m,n = 1,2,\ldots
 \label{zf}
\end{eqnarray}
In computing the matrix elements  $\langle m|d\rho|n\rangle$, we shall use the formula for the differentiation of an operator function \cite{W67} (for a prove in a rigorous mathematical setting, see also \cite{S97} and pp. 137-139 in ref. \cite{Z20})
\begin{equation}
\frac{\partial }{\partial h}e^{-H(h)}=-\int_0^1e^{-(1-u)H(h)}\frac{\partial H(h)}{\partial h}e^{-uH(h)}du,
\label{IW}
\end{equation}
where the operator $H(h)$ is a function of a parameter $h$.
The identity~\eqref{IW} is an indispensable ingredient of the theory when the family of density operators is of exponential form \cite{BT12,J14,MSP18,MA18,CSV18}.

For the one parameter family of Gibbs state, Eq.~\eqref{roh},  $h=0$, Eq.~\eqref{IW} immediately gives
\begin{equation}
\frac{\partial \rho (h)}{\partial h}\vert_{h=0}=\rho(0)\left[ \int_0^1 e^{T\lambda}Se^{-T\lambda} d\lambda- \langle S \rangle_T\right ],
\end{equation}
where
\begin{equation}
\langle \cdots \rangle_T:= [Z({T})]^{-1}\mathrm{Tr}\{e^{-T}\cdots\}
\end{equation}
denotes  the thermodynamic mean value.

 For the matrix elements of $d\rho$  (in the eigenbasis  of $\rho$), one obtains:
\begin{equation}
 |\langle m|d\rho|n\rangle|^2= |\langle m|\partial_h  H(h)|n\rangle|^2\frac {|\rho_n-\rho_m|^2}{|\ln\rho_n - \ln\rho_m|^2},\quad m\neq n,\quad \partial_h =\frac{\partial}{\partial h}
 \label{ur11}
\end{equation}
and
\begin{equation}
d\rho_m^2:=|\langle m|d\rho|m\rangle|^2=\rho_m[\langle m|S|m\rangle - \langle S \rangle_{T}].
\end{equation}

This result is a consequence of the exponential form of the density matrix and is particularly useful when $\rho$ is
known explicitly.
Plugging Eqs.~\eqref{ur11} and~\eqref{pr} in Eq.~\eqref{CMP} (and explicitly introduce in $d^2_f$ the dependence on S) one obtains
\begin{equation}
 d^2_{f}(S,S)\!=\!\frac{1}{4}\Big \{\langle (\delta S^d)^2\rangle_T +\!\!\!
\sum_{m,n,m\neq n}\!\!\!c_f(\rho_m,\rho_n)
\Big(\frac{\rho_n -\rho_m}{\ln \rho_n -\ln \rho_m}\Big)^2|\langle m| S|n\rangle |^2\Big \}.
\label{CMP1}
\end{equation}
In Eq.~\eqref{CMP1}, following \cite{ZVG07} (see also \cite{AASC10,BT12}) for the first term in Eq.(\ref{CMP}) we have used the relation
\begin{equation}
\frac{1}{4}\sum_{m}\frac{d \rho_m^2}{\rho_m}=\frac{1}{4}\langle (\delta S^d)^2\rangle_T =\sum_m \rho_{m}|\langle m|S|m \rangle|^2 - \langle S\rangle_T^2,
\label{vrz}
\end{equation}
where  $S^d:= \sum _{m}\langle m|S|m\rangle|m\rangle\langle m|$ is the diagonal part of the operator $S$ and  $\delta S^d := S^d - \langle S^d\rangle_T$.
The expression $d^2_{f}(S,S)$ constitutes Riemannian metrics parametrized by $f$ on the differentiable
manifold  $\mathcal{M}$.


 For our further consideration,
 it is useful  to introduce the family of functions announced in ref. \cite{T16}:
 \begin{equation}
g_f(x):=\frac{e^{2x}-1}{2xf(e^{2x})}\equiv \frac{1}{2x}\left [ \frac{1}{f(e^{-2x})}-\frac{1}{f(e^{2x})}\right]\ge 0, \quad f\in \mathcal{F}_{op}.
\label{gf}
\end{equation}
 Some examples of functions $g_{f}\left(x\right)$ for operator monotone functions  $f \in\mathcal{F}_{op}$ (see the Appendix) are the following:
\begin{alignat}{3}
&g_{Har}\left(x\right)=\frac{\sinh 2x}{2x},&\quad& g_{G}\left(x\right)=\frac{\sinh x}{x},&\quad& g_B\left(x\right)=\frac{\tanh x}{x}\,, \nonumber\\[-8pt]
\label{uga1}\\[-8pt]
&g_{BKM}\left(x\right)=1,&\quad&g_{WY}\left(x\right)=\frac{\tanh\frac{1}{2}x}{\frac{1}{2}x},&\quad& g_{MC}\left(x\right)=\frac{x}{\tanh x}\,.\nonumber
\end{alignat}
and
\begin{alignat}{2}
&g_{p}(x)=\frac{p}{1-p}\frac{1}{x}\sinh x
\frac{\sinh(p-1)x}{\sinh px}\,,&\quad& -1\le p \le 2,\nonumber\\[-8pt]
\label{thi}\\[-8pt]
&g^{\alpha}_{WYD}(x)=\frac{1}{2\alpha(\alpha-1)}\frac{1}{x}
\frac{\cosh x -\cosh(1-2\alpha)x}{\sinh x}\,,&\quad& 0\le\alpha\le 1.\notag
\end{alignat}
In general, the functions $g_f(x)$ satisfy  the condition
$g_f(x)=g_f(-x)$ and $g_f(0)= 1$.

Thus, Eq. (\ref{CMP1}) can be rewritten as:
\begin{equation}
 d^2_{f}(S,S)=\frac{1}{4}\Big \{\langle (\delta S^d)^2\rangle_T +
 \sum_{m,n,m\neq n}g_f\Big(\frac{1}{2}\ln\frac{\rho_n}{\rho_m}\Big)
\Big(\frac{\rho_n -\rho_m}{\ln \rho_n -\ln \rho_m}\Big)|\langle m| S|n\rangle |^2\Big \}.
\label{MSP}
\end{equation}
Using that by definition
\begin{equation}
\lim_{n \to m}\left\{g_f\left(\frac{1}{2}\ln\frac{\rho_n}{\rho_m}\right)
\left(\frac{\rho_n -\rho_m}{\ln \rho_n -\ln \rho_m}\right)\right\}=\rho_m,
\label{pr}
\end{equation}
Eq.~\eqref{MSP} may be rewritten in the form
\begin{equation}
d^2_{f}(S,S)=
\frac{1}{4}\Big \{\sum_{m,n,}g_f\Big(\frac{1}{2}\ln\frac{\rho_n}{\rho_m}\Big)
\Big(\frac{\rho_n -\rho_m}{\ln \rho_n -\ln \rho_m}\Big)|\langle m| S|n\rangle |^2 - \langle S\rangle_T^2\Big\},
\label{MSPr}
\end{equation}
which some time
is more convenient

 Replace $S$ with $\delta S:= S-\langle S \rangle _{T}$ in  Eq.~\eqref{MSPr}, since $ \langle \delta S \rangle_T \equiv 0 $,  one gets
\begin{equation}
d^2_{f}(\delta S,\delta S)=d^2_{f}(S,S),
\label{111}
\end{equation}
i.e. the replacement of $S$ by $S-\langle S \rangle_T$ does not alter the above definition of $d^2_f$. Therefore, when this does not cause confusion, for simplicity, we shall omit the dependence on $S$ in $d^2_f$.

Since $d^2_f$ collapse to $d^2_{BKM}(S,S)$ when $g_f(x)=g_{BKM}(x)=1$, Eq.~\eqref{MSP} (or equivalently Eq.~\eqref{MSPr}) prompts how to obtain a power series expansion of  $d^2_{f}(S,S)$ in terms on the moments of DSF \cite{T16}.

Finally, it is easily to obtain the inequalities
\begin{equation}
g_B\left(x\right)\le g_{WY}\left(x\right) \le g_{BKM}(x)\equiv 1\le g_{G}\left(x\right)\le  g_{MC}\left(x\right)\le g_{Har}\left(x\right)
\label{uga2}
\end{equation}
which imply inequalities between the different metrics $d_{f}(S,S)$.
Looking ahead let us notice that the symmetry relation $g_{B}(x)=g^{-1}_{MC}(x)$ provides an interesting inequality (see below)
\begin{equation}
\sqrt {d^2_{B}(S,S).d^2_{MC}(S,S)}\ge d^{2}_{BKM}(S,S).
\label{iei}
\end{equation}

The relation Eq. (\ref{MSP}) (or equivalently Eq.~\eqref{MSPr}) allows us to derive  linear response theory type sum-rules that may be useful to bound $d^2_{f}(S,S)$.

\section[]{Integral Presentation of $d^2_{f}(S,S)$ }\label{ts3}

After setting the relation
\begin{equation}
g_f\left(\frac{1}{2}\ln\frac{\rho_n}{\rho_m}\right) \equiv g_f\left(\frac{\omega_{nm}}{2}\right)
=
\int_{-\infty}^{\infty}g_f\left(\frac{\omega}{2}\right)\delta(\omega - \omega_{nm})d\omega,\quad
\omega_{nm}:=T_n - T_m,
\end{equation}
into Eq. (\ref{MSP}), one gets
\begin{equation}
d^2_{f}(S,S)=\frac{1}{4}\Big\{\langle (\delta S^d)^2\rangle_T
+
 \sum_{m,n,m\neq n}\int\limits_{-\infty}^{\infty}g_f\Big(\frac{\omega}{2}\Big)\delta(\omega - \omega_{nm})
 \rho_m\Big(\frac{1-e^{-\omega_{nm}}}{\omega_{nm}}\Big)|\langle m| S|n\rangle |^2d\omega\Big\}.
\label{dMSP}
\end{equation}

In the linear response theory  a main notion is the dynamical structure factor (DSF) \cite{K66,F80,ZMR96,BT15,PS04}
\begin{equation}
Q_{AB}(\omega)=[Z(T)]^{-1}\sum_{m,n}e^{-T_m}\langle n|A|m \rangle\langle m|B|n \rangle\delta(\omega -\omega_{nm}).
\label{ABsf}
\end{equation}
In Eq.~\eqref{ABsf}, in accordance with our initial convention the inverse temperature $\beta$  is absorbed  in the eigenvalues of the operators and we assume the Planck constant $\hbar$=1. We warm the reader that a definition of DSF differs from Eq.~\eqref{ABsf}.e.g.
\begin{equation}
\tilde{Q}_{AB}(\omega)=2\pi[Z(T)]^{-1}\sum_{m,n}e^{-T_n}\langle n|A|m \rangle\langle m|B|n\rangle\delta(\omega_{nm}-\omega).
\label{ABsf1}
\end{equation}
exists in the literature (see \cite{ZMR96}). If one uses that $T_n - T_m =\omega $ and hence
$e^{-T_m} =e^{-T_n} e^{\omega}$, due to the existence of the delta function in the summand of Eq.~\eqref{ABsf}, both definitions are related via  the relation
\begin{equation}
Q_{AB}(\omega)=\frac{1}{2\pi} e^{
	\omega}\tilde{Q}_{AB}(\omega).
\end{equation}
Notice that in case where the operator $B = A^{+}$, the  DSF, for all $T_{n}$,
is  real and positive definite.
In our case DSF is relative to the hermitian operator $S$ and we
shall use the notation:
\begin{equation}
Q_{S}(\omega)=[Z(T)]^{-1}\sum_{m,n}e^{-T_m}|\langle n|S|m \rangle|^2\delta(\omega -\omega_{nm}) \geq 0.
\label{sf}
\end{equation}

Now using Eq.~\eqref{sf} and the result for the thermodynamic  mean value of $(\delta S^d)^2$, Eq.~(\ref{vrz}), one may  recast  Eq.~(\ref{dMSP})  in the form
\begin{eqnarray}
d^2_{f}(S,S)=\frac{1}{4}\left\{\int_{-\infty}^{\infty}g_f\left(\frac{\omega}{2}\right)
 \left(\frac{1-e^{-\omega}}{\omega}\right)Q_{S}(\omega)d\omega - \langle S\rangle_T^2\right\}.
\label{pspMSP}
\end{eqnarray}
 Using
 the symmetry relation (named also detailed balancing relation)
\begin{equation}
Q_{S}(\omega)=Q_{S}(-\omega)e^{\omega}
\label{db}
\end{equation}
it is readily seen that the integrand in Eq.~(\ref{pspMSP}) is even non-negative function of $\omega$.
Thus,  the formula:
\begin{eqnarray}
d^2_{f}(S,S)=\frac{1}{4}\left\{2\int_{0}^{\infty}g_f\left(\frac{\omega}{2}\right)
 \left(\frac{1-e^{-\omega}}{\omega}\right)Q_{S}(\omega)d\omega - \langle S\rangle_T^2\right\}.
\label{spMSP}
\end{eqnarray}
is an alternative spectral  (or Lehmann) presentation of the  one to one correspondence between $d^2_{f}$ and the standard operator  monotone functions $f$.

Any choice  of the standard operator monotone function $f$ in  $g_f(x)$ generates a different metric.
Taking into account that the corresponding integrand in Eq.~\eqref{spMSP} is continuous and non-negative function for every $\omega \in [0,\infty]$, the inequalities \eqref{uga2}
state that
\begin{equation}
d^2_B\le d^2_{WY}\le d^2_{BKM}\le d^2_{G}\le  d^2_{MC}\le d^2_{Har}.
\label{uga3}
\end{equation}

\section[]{Presentation of $d^2_{f}(S,S)$ by Green's Functions}\label{ts4}

Before discussing the problem, let us  briefly recall  the
necessary background concerning the different presentations of the Green's functions \cite{ZMR96,BT15} we need. Here, we focus on the retarded and advanced Green's
function for any two  (non-Hermitian in general) operators $A(t)$ and $B(t^{'})$ given by
\begin{equation}
\langle\langle A,B\rangle\rangle^r(t)_{t}=-i\theta (t-t^{'})\langle[A(t),B(t^{'})]_{-}\rangle_H
\end{equation}
and
\begin{equation}
\langle\langle A,B\rangle\rangle^a_{t}=i\theta (t^{'}-t)\langle[A(t),B(t^{'})]_{-}\rangle_H
\end{equation}
respectively. Note that $A(t)$ and $B(t)$ are in the Heisenberg representation  referred to the Hamiltonian  of the system  $H$ we are interested in:
\begin{equation}
A(t)=e^{iHt}Ae^{-Ht},\quad B(t)=e^{iHt}Be^{-Ht}.
\label{t40}
\end{equation}
The Heaviside step function $\theta (t-t^{'})$ emerged as a natural consequence of causality.
 The Fourier transform of the retarded and advanced Green's functions
are given by
\begin{equation}
\langle\langle A,B\rangle\rangle^{r}_{\omega}=\langle\langle A,B\rangle\rangle_{\omega+i0^{+}}\label{ret}
={\cal P}\int_{-\infty}^{\infty}Q_{AB}(\omega^{'})(1-e^{-\omega})\frac{1}{\omega-\omega'}d\omega'-i\pi(1-e^{-\omega})Q_{AB}(\omega)
\end{equation}
and
\begin{equation}
\langle\langle A,B\rangle\rangle^{a}_{\omega}=\langle\langle A,B\rangle\rangle_{\omega-i0^{+}} \label{ad}
={\cal P}\int_{-\infty}^{\infty}Q_{AB}(\omega')(1-e^{-\omega})\frac{1}{\omega-\omega'}d\omega'+i\pi(1-e^{-\omega})Q_{AB}(\omega),
\end{equation}
respectively. The symbol ${\cal P}$ indicate that the principal value must be taken in the integrals.

One can obtain that $\langle\langle A,B\rangle\rangle_{E}^{r}$ satisfies  the algebraic equation
\begin{equation}
\omega \langle\langle A,B\rangle\rangle^{r}_{\omega}=\frac{i}{2\pi}\langle[A,B]\rangle + \langle\langle[C,H],B\rangle\rangle^{r}_{\omega},
\label{gfe}
\end{equation}
where $C=[A,H]$.

Let us introduce the complex-valued function
\begin{equation}
\langle\langle A,B\rangle\rangle _{E}=\int_{\infty}^{\infty}d\omega Q_{A,B}(\omega)\frac{1-e^{-\omega}}{E-\omega}.
\label{12c}
\end{equation}
The function $\langle\langle A,B\rangle\rangle_{E}$ is holomorphic
on the complex E-plane with cut along the real axis.We are now in position to apply the Bogoliubov and Tyablikov  spectral relation \cite{BT59}
\begin{equation}
(1-e^{-\omega})Q_{A,B}(\omega)=\frac{i}{2\pi}\{\langle\langle A,B\rangle\rangle_{\omega + i\epsilon} - \langle\langle \langle\langle A,B\rangle\rangle_{\omega - i\epsilon}\}
\label{BT}
 \end{equation}
in order to transfer the computational problem in Eq.~\eqref{pspMSP} in the realm of the Green's functions method.
For $A=B=\delta S$, where the product of the two matrix elements in $Q_{SS}(\omega)$ in the integrand of Eqs.~\eqref{ret} and~\eqref{ad} is real, the function $\langle\langle A,B\rangle\rangle_{\omega + i\epsilon}$ is the complex
conjugate of the  function $\langle\langle A,B\rangle\rangle_{\omega - i\epsilon}$.
In this particular case,
\begin{equation}
\pi(1-e^{-\omega})Q_{\delta S\delta
	S}(\omega)=-\mathrm{Im} \langle\langle S,S\rangle\rangle ^{r}_{\omega},
\label{BT1}
\end{equation}
(here $\omega$ is a real quantity) and after plugging Eq.~\eqref{BT1} in  Eq.~\eqref{pspMSP}
the following relation holds:
\begin{eqnarray}
d^2_{f}(\delta S,\delta S)=-\frac{1}{4}\left\{\frac{1}{\pi}\int_{-\infty}^{\infty}\frac {1}{\omega}g_f\left(\frac{\omega}{2}\right)
\mathrm{Im} \langle\langle S,S\rangle\rangle ^{r}_{\omega} d\omega \right \}.
\label{p4MSP1}
\end{eqnarray}

The explicit determination of the Green's functions or, equivalently, of the DSF generally requires the full solution of the
infinite chain of equations, Eq.~\eqref{gfe},
or the solution of the Schr\"{o}dinger equation, yielding the eigenvalues of the Hamiltonian and matrix elements of Eq.~\eqref{sf}. In most interacting systems, the solution of Eqs. (\ref{gfe}) is a difficult task that can usually  be
accomplished only approximately.

We show that monotone Riemannian metrics can be determined from an analysis
of the equation of motion of the retarded Green's function. An alternative way is based on the analysis of Green's functions using Feynman diagrams.

Let us recall the relation \cite{ZMR96,PS04}:
\begin{equation}
\chi^{''}_{S}(\omega)
=-\mathrm{Im}\langle\langle S,S\rangle\rangle _{\omega}^r,
\label{chi}
\end{equation}
where $\chi^{''}_{S}(\omega)$ is the dissipative component  of the Kubo response  function.
Combating Eq.~\eqref{BT1} and Eq.~\eqref{chi} one obtains the fluctuation dissipation theorem
\begin{equation}
\chi^{''}_{S}(\omega)=\pi(1-e^{-\omega})Q_{\delta S\delta
	S}(\omega).
\label{FDT}
\end{equation}
Thus Eq. (\ref{p4MSP1})  may be presented as
\begin{equation}
d^2_f(S,S)=\frac{1}{4}\left\{\frac{1}{\pi}\int_{-\infty}^{\infty} \frac{1}{\omega}g_{f}\left(\frac{\omega}{2}\right)\chi^{''}_{S}(\omega)d\omega\right\}.
\label{MRM1c}
\end{equation}
Here, it is useful to recall that $\chi^{''}_{S}(\omega)$ is a real odd
function.

\section[]{Presentation of $d^2_f(S,S)$ within a Thermodynamic Setting and\\ Some Particular Inequalities}\label{ts5}

In order to establish the neat relation of Eq.~\eqref{MRM1c} with some  thermodynamic quantities and inequalities we shall consider the following  important particular cases:

\subsection[]{The Bogoliubov-Kubo-Mori function $f_{BKM}(x)=\frac{x-1}{\ln x}$}\label{ts51}

Let us define
 the Bogoliubov-Duhamel  inner product \cite{B61,DLS,PT,S93,R09,BT11,P94} (which is often called Bogoliubov-Kubo-Mori scalar product or canonical correlation) for the operators  $A$ and $B$  by the formula:
 \begin{equation}
 F_{0}(A;B):=\int_0^1d\tau \left\langle e^{\tau T} A^+ e^{-\tau T}B\right\rangle_T
 \label{vaz}
 =\frac{1}{2}\sum_{m,n, m\not=n} |\langle n|A^+|m \rangle \langle n|B|m \rangle|\frac{\rho_n
 	-\rho_m}{X_{mn}}+ \sum_{n}\rho_n
 |\langle n|A^+|n \rangle \langle n|B|n \rangle|.
 \end{equation}
  Note that $ F_0( \delta S;\delta S)$ is an important ingredient of the linear  response theory, and is exactly the isothermal susceptibility associated with $h$, $\chi_{(h=0)}$
 \begin{equation}
 \chi_{(h=0)}=F_{0}(\delta S;\delta S)=\frac{1}{\pi}\int_{-\infty}^{\infty} \omega^{-1}\chi^{''}_{S}(\omega)d\omega.
 \label{MRM3c}
 \end{equation}
In this case $g_{BKM}(\omega/2)=1$. From Eq. (\ref{pspMSP})  we obtain  the well known result:
\begin{eqnarray}
d^2_{BKM}(S,S)=\frac{1}{4}F_{0}(\delta S;\delta S),\quad \delta S = S - \langle S\rangle_T.
\label{dBKM}
\end{eqnarray}
From the other side, from Eq.~\eqref{MRM1c} one gets
\begin{equation}
d^2_{BKM}(S,S)
=\frac{1}{8\pi}\int_{-\infty}^{\infty} \left(\frac{\omega}{2}\right)^{-1}\chi^{''}_{S}(\omega)d\omega=\frac{1}{4}\chi_{(h=0)}.
\label{MRM2c}
\end{equation}

Using the idea of so called generalized or deformed metrics, the physical interpretation of the Bogoliubov-Kubo-Mori metric as an integral (global) characteristic of the one-parameter family of Wigner-Yanase-Dyson metrics was clarified and its intermediate position between extremal metrics was analyzed in ref.~\cite{R09}. For more details one can see also ref.~\cite{R13}.

\subsection[]{The   Morozova-$\breve{C}$encov  function $f_{MC}(x)=\left(\frac{x-1}{\ln x}\right)^2\frac{2}{1+x}$}\label{ts52}

Note that the subscript  $MC$  means that the operator monotone function $f_{MC}$ was introduced by  Morozova and $\breve{C}$encov \cite{MC90}.
In this case, plugging   $g_{MC}(\frac{\omega}{2})=\frac{\omega/2}{\tanh \omega/2}$ in  Eq.~\eqref{pspMSP} with the help of the identity
\begin{equation}
\tanh \frac{\omega}{2}= \frac{1-e^{-\omega}}{1+e^{-\omega}}
\end{equation}
and Eq.~\eqref{db} we obtain \cite{T16}:
\begin{equation}
d^2_{MC}(S,S)
=\frac{1}{4}{\langle(S-\langle S \rangle_T)^2\rangle_T}.
\label{dMC}
\end{equation}
From the other side, combining Eq.~(\ref{MRM1c}) with Eq.~(\ref{dMC})   one gets the relation:
\begin{equation}
\langle S^2-\langle S \rangle_T^2 \rangle_T =\frac{1}{2\pi}\int_{-\infty}^{\infty}\chi^{''}_{S}(\omega)\coth \left(\frac{\omega}{2}\right)d\omega
\label{MRM1d}
\end{equation}
also known as the Callen-Welton fluctuation-dissipation theorem (see, e.g. \cite{ZMR96,PS04}).

If operators $T$ and $S$ commute, one can see  from Eq.~(\ref{vaz})
that $d_{BKM}^2$ and $d_{MC}^2$ coincides.
It is instructive to consider
 the difference between the total fluctuations of an observable $\langle S^2-\langle S \rangle_T^2 \rangle_T$ and its thermal fluctuations $F_{0}(\delta S;\delta S)$  studied in refs. \cite{MA18,F18,FR16,FR19}. This difference called ``quantum variance'' is studied in \cite{FR16} as an important  quantity that  provides a tight lower bound to some most widely accepted estimators of ``quantumness''.  In ref.~\cite{MA18}, in a slightly modified version of the exponential model given by Eq.~\eqref{roh}, it was
 demonstrated that the strictly classical
 fluctuations in $S$ constrain the achievable precision in
 estimates of $h$.

 In the context of the Riemannian metrics  the corresponding expression (in our notation) is \cite{BT11, FR19}:
\begin{equation}
d^2_{MC}(S,S)-d^2_{BKM}(S,S)\!=\!\frac{1}{8\pi}\int\limits_{-\infty}^{\infty}\chi{''}_{S}(\omega)
\Big(\frac{\omega}{2}\Big)^{-1}\Big[\Big(\frac{\omega}{2}\Big)\coth \Big(\frac{\omega}{2}\Big)-1\Big]d\omega.
\label{FR16}
\end{equation}
The above equation is a particular case of $f= f_{MC}$ of the more general relation
\begin{equation}
d^2_{f}(S,S)-d^2_{BKM}(S,S)=\frac{1}{8\pi}\int_{-\infty}^{\infty}\chi^{''}_{S}(\omega)\left(\frac{\omega}{2}\right)^{-1}\left[g_f \left(\frac{\omega}{2}\right)-1\right]d\omega.
\label{MG}
\end{equation}
The explicit determination of the rhs of Eq.~\eqref{MG} requires the knowledge of $\chi^{''}_{S}(\omega)$ which in itself is a difficult task. It is shown \cite{F18,FR16,FR19} that the rhs of Eq.~\eqref{FR16} lends itself to analysis based on the Feynman path-integral representation as well to numerical Monte-Carlo  based analysis.

It is worth noting that
 in ref.~\cite{T16} the deviation  of any monotone
Riemannian meric $d_f^2(S,S)$ from $d^2_{BKM}(S,S)$ (or $d^2_{MC}(S,S)$)
has been presented as a series expansion
in terms of the moments of DSF relative to the operator $S$ (see also below). 
 An useful information may be obtained by
 some thermodynamic inequalities as well. Different choices of the upper bound on the rhs of Eq.~\eqref{FR16} may generate different thermodynamic inequalities \cite {BT11,BT13}.
 For example the application of the elementary inequality
\begin{equation}
1\le x\coth x \le 1 + \frac{1}{3}x^2
\end{equation}
to the rhs of Eq.~\eqref{FR16}  immediately yields
\begin{equation}\label{BHFR16}
0\le d^2_{MC}(S,S)-d^2_{BKM}(S,S)
\le\frac{1}{48\pi}\int_{-\infty}^{\infty}\chi{''}_{S}(\omega) \omega d\omega
=\frac{1}{48}\langle [[S,T]_{-},S]_{-} \rangle_T,
\end{equation}
which is a reminiscent of the well known  thermodynamic inequality of Brooks Harris \cite{BH67}  (see also \cite{BT11}).

\subsection[]{The Bures function $f_B(x) = \frac{x+1}{2}$}\label{ts53}


In this case $g_B(\omega/2)=\dfrac{\tanh \omega/2}{\omega/2}$, and one has the expression (named also Bures metric or fidelity susceptibility, see e.g. \cite{BT12,TB13,T14,J14}):
\begin{eqnarray}
d^2_{B}(S,S)=\frac{1}{2}\left\{\int_{-\infty}^{\infty}\frac{\tanh (\omega/2)}{\omega}
\left(\frac{1-e^{-\omega}}{\omega}\right)Q_{\delta S}(\omega)d\omega \right\}.
\label{1pspMSP}
\end{eqnarray}
With the help of the fluctuation-dissipation theorem, Eq.~\eqref{FDT},
from Eq.~\eqref{1pspMSP} one gets
\begin{equation}
d^2_{B}(S,S)=\frac{1}{8\pi}\int_{-\infty}^{\infty}\left(\frac{\omega}{2}\right)^{-2}\tanh \left(\frac{\omega}{2}\right)\chi^{''}_{S}(\omega)
d\omega,
\label{1ppMSP}
\end{equation}
(see also \cite{CSDV19}).

A similar expression to Eq.~\eqref{FR16} holds for  $d^2_{B}(S,S)$ instead of $d^2_{MC}(S,S)$:
\begin{equation}
d^2_{BKM}(S,S)-d^2_{B}(S,S)=\frac{1}{8\pi}\int_{-\infty}^{\infty}\chi{''}_{S}(\omega)\left(\frac{\omega}{2}\right)^{-1}\left[1-\left(\frac{\omega}{2}\right)^{-1}\coth^{-1} \left(\frac{\omega}{2}\right) \right]d\omega.
\label{TFR16}
\end{equation}
An alternative presentation of $d^2_{B}(S,S)$ in terms of the thermodynamic mean values of successively   higher commutators of the Hamiltonian  with the operator involved  through the control parameter (e.g. $h$) is given in \cite{T14}:
\begin{equation}
d_{BKM}^2(S,S)-d^2_{B}(S,S)=2\sum_{l=1}^{\infty}
\frac{2^{2l+2}}{(2l+2)!}B_{2l+2}\langle R_{2l-1}R_0\rangle_T,
\label{Ton}
\end{equation}
where the iterated commutators (see, also subsection~\ref{ts71} bellow) $R_n\equiv R_n(S)=[T,R_{n-1}(S)],\quad n=1,2,...$ and $B_{2n}$ are the Bernoulli numbers.

The equivalence  of  Eq.~\eqref{Ton} and Eq.~\eqref{TFR16} may be obtained with the help of the
the relation
\begin{equation}
\langle R_{2l-1}R_0\rangle_T=-M_{2l-1}(S),
\end{equation}
where $M_{2l-1}(S)$ are the moments of the DSF (see Eq.~\eqref{Mchi} 
in the next subsection~\ref{ts72}).

From Eq.~\eqref{TFR16}, with the help of the elementary inequality
\begin{equation}
1-\frac{1}{3}x^2\le (x\coth x)^{-1} \le 1,
\end{equation}
 immediately follows
\begin{equation}
0\le d^2_{BKM}(S,S)-d^2_{B}(S,S)\le \frac{1}{48\pi}\int\limits_{-\infty}^{\infty}\chi^{''}_{S}(\omega) \omega d\omega =\frac{1}{48}\langle [[S,T]_{-},S]_{-} \rangle_T.
\label{R16}
\end{equation}
The above inequalities first were proven in \cite{BT12} in terms of the fidelity susceptibility, where their usefulness is illustrated by two examples: the Dicke model of superradiance and sigle-impurity Kondo model.
The lhs of \eqref{R16} was proven also in \cite{MA18}) and used to to obtain a new uncertainty relation between energy and
temperature for a quantum system strongly interacting
with a reservoir.

Finally, for the difference
\begin{equation}
 d^2_{MC}(S,S)-d^2_{B}(S,S)= \frac{1}{8\pi}\int\limits_{-\infty}^{\infty}\chi{''}_{S}(\omega)\Big(\frac{\omega}{2}\Big)^{-1}\Big[
\frac{\omega/2}{\tanh(\omega/2)}-\frac{\tanh\left(\omega/2\right)}{\omega/2}\Big]d\omega.
\label{NTi}
\end{equation}
with the help of the inequality
\begin{equation}
0\le
\frac{x}{\tanh x}-\frac{\tanh x}{x}\le\frac{2}{3}x^2
\end{equation}
 one obtains the inequality
\begin{equation}
0\le \le d^2_{MC}(S,S)-d^2_{B}(S,S)\le\frac{1}{24}\langle [[S,T],S], \rangle_T.
\label{nNiT}
\end{equation}

If the Brook Harris inequality \eqref{BHFR16} imposes a restriction on the isothermal susceptibility $ \chi_{h=0}\equiv 4 d^2_{BKM} (S,S)$ then \eqref{nNiT} is its counterpart for the fidelity susceptibility $ \chi_{F}\equiv 4 d^2_{B} (S,S)$.

\subsection[]{The Wigner-Yanase-Dayson function}\label{ts54}

 The Wigner, Yanase and  Dyson (WYD) skew information \cite{WY63} (see also \cite{FY12} and refs. therein) is given by
\begin{equation}
I^{{f_{WYD}}}(\rho,S):=-\frac{1}{2}[Z({T})]^{-1}\text{Tr}\left([e^{-\alpha T},S^+]_{-}\times[e^{-(1-\alpha)T},S]_{-}\right),\quad
0 \leq \alpha \leq 1.
\label{wyd0}
\end{equation}
Here, we use the  superscript $f_{WYD}$ to stress that $I^{f_{WYD}}(\rho,S)$ is related to
the standard operator monotone function
\begin{equation}
f_{WYD}\equiv f_{WYD}(\alpha,x) = \alpha(\alpha-1)\frac{(x-1)^2}{(x^\alpha-1)(x^{1-\alpha}-1)},\quad 0<\alpha<1.
\label{WYD1}
\end{equation}
 The WYD-skew information
 can also be written as
 \begin{equation}
 I^{{f_{WYD}}}(\rho,S)=\text{Tr}\{\rho S^2\}-\text{Tr}\{\rho^{\alpha}S\rho^{1-\alpha}S\}.	
 	\end{equation}
One may observe that
\begin{equation}
\frac{1}{4}\int_0^1I^{{f_{WYD}}}(\rho,S)d\alpha=d^2_{MC}(S,S)-d^2_{BKM}(S,S) \le
\frac{1}{48\pi}\int_{-\infty}^{\infty}\chi^{''}_{S}(\omega) \omega d\omega=\frac{1}{48}\langle [[S,T]_{-},S]_{-}, \rangle_T,
\label{R09}
\end{equation}
and thus (see Eq.~\eqref{BHFR16})
\begin{equation}
\int_0^1I^{{f_{WYD}}}(\rho,S)d\alpha \le\frac{1}{12\pi}\int_{-\infty}^{\infty}\chi^{''}_{S}(\omega) \omega d\omega=\frac{1}{48}\langle [[S,T]_{-},S]_{-}, \rangle_T,
\label{R09a}
\end{equation}

A generalization of Eq.~\eqref{wyd0} (introduced  in ref. \cite{H08} as a ``metric adjusted skew information'') is given  (in our notations) by
\begin{equation}
I^f(\rho,S)=\frac{f(0)}{2}\sum_{m,n}\frac{(\rho_m-\rho_n)^2}{\rho_nf(\rho_m/\rho_n
	)}|
\langle m|S| n\rangle|^2,
\label{aSI}
\end{equation}
where $f \in\mathcal{F}_{op}$ is an arbitrary standard operator monotone function  with  $f(0)\neq 0$, see also \cite {PG11}. Plugging
Eq.~\eqref{WYD1} in Eq.~\eqref{aSI} one obtains as a particular case the WYD-skew information given by Eq.~\eqref{wyd0}.

Here,  we shall consider a set of Gibbs states characterized by the  family of  density matrices given by
\begin{equation}
\rho(h) = [Z_N(h)]^{-1}\exp[- H(h,R_1)], \label{roh1}
\end{equation}
defined on the family of $N$-particles Hamiltonians of the form
\begin{equation}
H(h)= T - h R_1,\qquad R_1:=[T,S]_{-}
\label{ham}
\end{equation}
where the Hermitian operators $T$ and $S$ do not commute.
Formally, replacing  $S$ by $R_1$  in Eq.~\eqref{MSPr} after a straightforward calculation one obtains
\begin{equation}
\tilde {d}^2_{f}(S,S):=d^2_{f}(R_1,R_1)=
\frac{1}{4}\left \{\sum_{m,n,}g_f\left(\frac{1}{2}\ln\frac{\rho_n}{\rho_m}\right)
\left(\frac{\rho_n -\rho_m}{\ln \rho_n -\ln \rho_m}\right)\left(\ln \frac{\rho_n}{\rho_m}\right)^2|\langle m| S|n\rangle |^2 \right\},
\label{MSPr1}
\end{equation}
where the symbol ``tilde'' is used to emphasize the change of the statistical model, i.e. Eq.~\eqref{roh1} instead of Eq.~\eqref{roh}.
Comparing  Eq.~\eqref{aSI} and Eq.~\eqref{MSPr1} one obtains
\begin{equation}
\frac{1}{4}I^f(\rho,S)=\frac{f(0)}{2}\tilde {d}^2_{f}(S,S).
\label{FH}
\end{equation}
Recall that the same relation (up to the irrelevant multiplier $1/4$) by definition holds between WYD-skew information and QFI (see Definition 1.2 in \cite{H08}).

If one strictly follows the reasoning of Section~\ref{ts4}, the following result for $\tilde{d}^2_f(S,S)$ takes place
\begin{equation}
\tilde{d}^2_f(S,S)=\frac{1}{4}\left\{\frac{1}{\pi}\int_{-\infty}^{\infty} \omega g_{f}\left(\frac{\omega}{2}\right)\tilde{\chi}^{''}_{S}(\omega)d\omega\right\}.
\label{MRM12}
\end{equation}
Plugging  Eq.~\eqref{MRM12} in
Eq.~\eqref{FH} one gets
\begin{equation}
I^f(\rho,S)=\frac{f(0)}{2\pi}\int_{-\infty}^{\infty} \omega g_{f}\left(\frac{\omega}{2}\right)\tilde{\chi}^{''}_{S}(\omega)d\omega.
\label{ISK}
\end{equation}
Formally, this result coincides with Eq.~(40) of ref.~\cite{SU16}. The difference is hidden in the physical meaning of  $\tilde{\chi}^{''}_{S}$ where the external perturbation of a special type, i.e $R_1$ is applied.

If we consider the standard operator monotone function given by Eq.~\eqref{WYD1} then the metric adjusted skew information is
\begin{equation}
I^{f_{WYD}}(\rho,S)=\frac{1}{2}\left\{\frac{1}{\pi}\int_{-\infty}^{\infty}
\frac{\cosh(\omega/2) -\cosh[(1-2\alpha)\omega/2]}{\sinh (\omega/2)}\chi^{''}_{S}(\omega)d\omega\right\},
\end{equation}
or using Eq.~\eqref{MRM1d}
\begin{equation}
I^{f_{WYD}}(\rho,S)=\langle S^2-\langle S \rangle_T^2 \rangle_T+
\frac{1}{2}\left\{\frac{1}{\pi}\int_{-\infty}^{\infty}
\frac{\cosh[(1-2\alpha)\omega/2]}{\sinh (\omega/2)}\chi^{''}_{S}(\omega)d\omega\right\}.
\end{equation}
Integrating the above equation over $\alpha$ one obtains (with the help of Eq.~\eqref{dBKM}) the well known result for the WYD-skew information
\begin{equation}
\int_0^1I^{f_{WYD}}(\rho,S)d\alpha=\langle S^2-\langle S \rangle_T^2 \rangle_T-F_0((\delta S;\delta S)),
\end{equation}
or for the metrics
\begin{equation}
\frac{1}{8}\int_0^1\tilde {d}^2_{f_{WYD}}(S,S)d\alpha=d_{MC}(S,S) - d_{BKM}(S,S).
\end{equation}


\section[]{Applications of the Integral Presentation: General Inequalities}\label{ts6}


For any two  (non-Hermitian in general) operators $\delta A = A - \langle A \rangle_T$ and $B=A - \langle B \rangle_T$ let us define
\begin{equation}
P_{A;B}(\omega)=(1+e^{-\omega})Q_{\delta A,\delta B}(\omega),
\end{equation}
where
\begin{equation}
Q_{\delta A,\delta B}(\omega)=[Z(T)]^{-1}\sum_{m,n}e^{- T_m}\langle|n|\delta A|m \rangle)(m|\delta B|n\rangle\delta(\omega -\omega_{nm}).
\label{sfAB}
\end{equation}
Thus a straightforward generalization of our formula (\ref{pspMSP}) is:
\begin{equation}
d^2_{f}(\delta A,\delta B)=\frac{1}{4}\left\{\frac{1}{2}\int_{-\infty}^{\infty}\left(\frac{\omega}{2}\right)^{-1}\tanh \left(\frac{\omega}{2}\right)g_{f}\left(\frac{\omega}{2}\right)P_{ A, B}(\omega)d\omega \right\}.
\label{DAB}
\end{equation}
It is easy to check that $P_{\delta A,\delta B}(\omega)$ defines a scalar product
in the space of operators $\delta A$ and $\delta B$ and satisfies  the Cauchy-Schwartz  inequality in the form (at this point we utilize  the idea of ref. \cite{S92})
\begin{equation}
\Big|\int_{-\infty}^{\infty}G_A^f(\omega)G_B^{\tilde f}(\omega)P_{A;B}(\omega)d\omega\Big|^2\leq \int_{-\infty}^{\infty}|G_A^f(\omega)|^2P_{A;A}(\omega)d\omega
\int_{-\infty}^{\infty}|G_{B}^{\tilde f}(\omega)|^2P_{B;B}(\omega)d\omega,
\label{CSc}
\end{equation}
for any two functions (complex in general) $G_A^{f}(\omega)$ and $G_B^{\tilde f}(\omega)$  labeled by the operators $A$ and $B$ and the standard operator  monotone functions $f(\omega)$ and $\tilde {f} (\omega)$.

	A.) Let us define
	\begin{equation}
	\tilde f(x)=\sqrt{f(x)\bar f(x)},
	\label{fr}
	\end{equation}
where $\tilde f(x),f(x)$ and $\bar f(x)$ are standard operator monotone functions.
	By setting
\begin{eqnarray}
&&G_A^{f}(\omega)=\left(\frac{\omega}{2}\right)^{-1/2}\left[\tanh \left(\frac{\omega}{2}\right)g_{f}(\omega/2)\right]^{1/2},\nonumber\\
&& G_B^{\tilde f}(\omega)=\left(\frac{\omega}{2}\right)^{-1/2}\left[\tanh \left(\frac{\omega}{2}\right)g_{^{\tilde f}}(\omega/2)\right]^{1/2}
\end{eqnarray}
in \eqref{CSc}, one gets the inequality
\begin{multline}
\left |\int_{-\infty}^{\infty}\left(\frac{\omega}{2}\right)\left[\tanh \left(\frac{\omega}{2}\right)g_{\tilde{f}} (\omega/2)\right]P_{A;B}(\omega)d\omega\right |^2\\
\le\int_{-\infty}^{\infty}\left(\frac{\omega}{2}\right)\left[\tanh \left(\frac{\omega}{2}\right)g_{f}(\omega/2)\right]P_{A;A}(\omega)d\omega\\
\times
\int_{-\infty}^{\infty}\left(\frac{\omega}{2}\right)\left[\tanh \left(\frac{\omega}{2}\right)g_{\bar{f}} (\omega/2)\right]P_{B;B}(\omega)d\omega.
\label{mcsh}
\end{multline}
Thus, using Eq.~\eqref{DAB},
 one obtains
\begin{equation}
|d^2_ {\tilde f}(\delta A,\delta B)|^2\leq d^2_{f}(\delta A,\delta A)d^2_{\bar f}(\delta B,\delta B),
\label{ner}
\end{equation}
 It is readily seen that if
 $A=B=S$ and all three functions $\bar {f}(x),\tilde {f}(x)$ and $f(x)$ are {\it standard operator monotone functions} obeying the functional relation Eq.~\eqref{fr}, inequality (\ref{ner}) generates inequalities between Riemannian metrics. The geometric mean of two  operator monotone functions is still an operator
 monotone function in the following important cases:
 \begin{equation}
 f_{BKM}(x)=\sqrt{f_B(x)f_{MC}(x)}, \quad f_{G}(x)=\sqrt{f_B(x)f_{Har}(x)},
 \end{equation}
 where $f_{G}(x):=\sqrt{x}$, and thus the inequalities are true:
 \begin{equation}
 d_{G}^2 \leq ( d_{B}^2 \cdot d_{Har}^2)^{1/2}
\label{in2}
 \end{equation}
 and
\begin{equation}
d_{BKM}^2 \leq ( d_{B}^2 \cdot d_{MC}^2)^{1/2}.
\label{in1}
\end{equation}
Let us note that inequalities (\ref{in2}) and (\ref{in1}) are stronger than $d_{G}^2 \leq d_{Har}^2$ and $d^2_{BKM}\leq d^2_{MC}$, respectively.

	B.)
	We do remark that a particular realization of the Eq.~\eqref{fr} takes place for the both monotone operator functions 	$f_{\frac{1}{2}-d}(x)$ and $f_{\frac{1}{2}+d}(x)$ defined by Eq.~\eqref{clmf}, and $f_{G}(x)$ \cite{F08}.
		Then
	it is easily verified that
	from Eq.~\eqref{thi} one gets the relation
	\begin{equation}
	g_{\frac{1}{2}-d}(x)g_{\frac{1}{2}+d}(x)=\frac{\sinh^2(x)}{x^2}\equiv [g_G(x)]^2.
	\end{equation}
As a result the Cauchy-Schwartz  inequality \eqref{mcsh} implies the inequality
	\begin{equation}
 d_{G}^2 \leq \left( d_{\frac{1}{2}-d}^2 \cdot d_{\frac{1}{2}+d}^2\right)^{1/2},\quad 0 \le d \le \frac{3}{2}.
	\end{equation}

	C.) By setting
\begin{equation}
G_A(\omega)=\left(\frac{\omega}{2}\right)^{-1}\tanh \left(\frac{\omega}{2}\right)[g_{f}(\omega/2)]^{1/2},\quad G_B(\omega)= [g_{f}(\omega/2)]^{1/2},
\end{equation}
one gets
\begin{equation}
|d^2_{f}(\delta A,\delta B)|^2\leq\int_{-\infty}^{\infty}[\left(\omega/2\right)^{-2}\tanh^2 \left(\omega/2\right)g_{f}(\omega/2)] P_{A;A}(\omega)d\omega
\int_{-\infty}^{\infty}[g_{f}(\omega/2)]P_{B;B}(\omega)d\omega,
\end{equation}
or if $A=B=S$ and $f=f_{BKM}$ one gets
\begin{equation}
d_{BKM}^2 \leq ( d_{B}^2 \cdot d_{MC}^2)^{1/2},
\label{in1a}
\end{equation}
or if $A=B=S$ and $f=f_{B}$ one gets
\begin{equation}
d_{B}^2 \leq (d_{BKM}^2)^{1/2}\left\{\int_{-\infty}^{\infty}\left[\frac{\tanh(\omega/2)}{\omega/2}\right]P_{B;B}(\omega)d\omega \right\}^{1/2},
\label{in1b}
\end{equation}

The application of the inequality  $(x\coth x)^{-1}\le 1$ to Eq.~\eqref{DAB} readily gives the following  upper bound:
\begin{equation}
 d^2_{f}(\delta A,\delta B)\le\frac{1}{4}\left\{\frac{1}{2}\int_{-\infty}^{\infty}g_{f}\left(\frac{\omega}{2}\right)P_{ A, B}(\omega)d\omega \right\},
\label{DA1B}
\end{equation}
which becomes an identity in the classical regime of high temperature
 where $\coth(\omega/2) \to \omega/2$ in \eqref{DAB} (recall that the inverse temperature is absorbed in $\omega$).

In the particular case of Bures metric, $g_B\left(x\right)=\omega^{-1}\tanh x$, from the above inequality one gets \cite{BT12}
\begin{equation}
d^2_{B}(\delta A,\delta B)\le\frac{1}{4}\left\{\frac{1}{2}\int_{-\infty}^{\infty}g_{B}\left(\frac{\omega}{2}\right)P_{ A, B}(\omega)d\omega \right\}=d^2_{BKM}(\delta A,\delta B),
\label{DA21B}
\end{equation}
in consistent also with~\eqref{in1b}.

\section[]{Monotone Riemannian Metrics (Quantum Fisher Informations)\\ within the Set Defined by Ref.~\cite{T16}}\label{ts7}

In this Section, first we shall review how ref.~\cite{T16}  relates the entire class of MRM to the moments of DSF.
Thus, useful information of their behavior  is given in terms of the thermodynamic  mean values of iterated commutators of the considered Hamiltonian $T$ with the operator $S$ involved through  the control parameter $h$. After reviewing the result of
ref.~\cite{T16}, we shall connect the results of present study to that of \cite{T16}.

\subsection{Summary of the basic results of ref.~\cite{T16}}\label{ts71}
Let us introduce the moments of the  DSF:
\begin{equation}
M_{p}(S):=\int_{-\infty}^{+\infty}d\omega \omega^{p}Q_{S}(\omega),\quad p=-1,0,1,2,...
\label{dm11}.
\end{equation}
It is  possible to check that
\begin{equation}
M_{p-1}(S)=2^{-1}F_{p}(S;S),\quad p=0,1,2,...
\label{VRNTa}
\end{equation}
(for details  see \cite{T16}), where the functionals $F_n(S;S):$
\begin{equation}
\begin{split}
&F_{p}(S;S) := 2^{n-1} \sum_{ml}|\langle m|S|l \rangle |^{2}|\rho_{l}
- (-1)^p \rho_{m}| \cdot |X_{ml}|^{n-1},\\
&X_{ml}:= 2^{-1}  (T_m-T_l),\quad  p=0,1,2,...
\end{split}
\label{NT10ab}
\end{equation}
have been introduced in \cite{BT11}
as a generalization of the Bogoliubov - Duhamel inner product $F_0(S;S)$, see above the Eq.~(\ref{vaz}).
It is convenient  to use the following  basis independent presentations of  $F_{n}(S;S)$ \cite{T14,T16}:
\begin{equation}
F_{p}(S;S)=2(-1)^{p+1}\langle R_{p-1}R_{0}\rangle_T,\quad
p=0,1,2,....
\label{Nf1},
\end{equation}
where the notion of iterated  commutators (named also nested commutators)
\begin{eqnarray}
&&R_{0}\equiv R_{0}(S)\equiv S,\quad  R_{1}\equiv R_{1}(S) :=[T,S],\;
\dots, \; \nonumber\\
&& R_{p} \equiv R_{p}(S):= [T,R_{p-1}(S)]_{-},\quad p=0,1,2,...,
\label{R1a}
\end{eqnarray}
is introduced. By definition
$R_{-1}\equiv X_{ST}$ is a solution of the operator equation
\begin{equation}
S = [T,X_{ST}]_{-}
\label{oe}.
\end{equation}
Eqs.(\ref{VRNTa})  are not but  the well known sum rules for the moments of the DSF in the linear response theory (see, e.g. \cite{K66,F80,ZMR96,PS04}).
Note that Eqs.(\ref{VRNTa})  provide an algebraic way to evaluate the moments of the DSF.

{\it Main results.} The main results concern two series expansions of $d^2_f$ which quantify the deviation of any one member of the family $d^2_f$ from $d^2_{BKM}$ or $d^2_{MC}$, respectively.

The following formulas are valid \cite{T16}:

A.)
\begin{equation}
d^2_{f}(S,S)=  \left\{d^2_{BKM} +  \frac{1}{4}\sum_{l=1}^{\infty}\left(\frac{1}{2}\right)^{2l-1} a_{2l-1}(f) M_{2l-1}(S)\right\},
\label{1MSPna}
\end{equation}
where  the formal series expansion of the family of  functions $g_f(x)$:
\begin{equation}
g_{f}\left(x\right)=1 + \sum_{l=1}^{\infty}a_{2l-1}(f) (x)^{2l}
\label{use}
\end{equation}
defines the infinite sequence of coefficients (which depend on $f$)  $a_{2l-1}(f)$,$~l=1,2,...$ in the series expansion in Eq.~(\ref{1MSPna}).

B.)

\begin{equation}
d^2_{f}(S,S)=\left\{d^2_{MC} + \frac{1}{4}\sum_{l=1}^{\infty} \left(\frac{1}{2}\right)^{2l}a_{2l}(f) M_{2l}(S)\right \}.
\label{1MSPnb}
\end{equation}
where the formal series expansion
\begin{equation}
\hat{g}_{f}\left(x\right)=1 + \sum_{l=1}^{\infty}a_{2l}(f) (x)^{2l},
\label{use1}
\end{equation}
of the family of generating functions
\begin{equation}
\hat{g}_f\left(x\right):=g_f\left(x\right)\frac{\tanh x}{x}
\label{hat}
\end{equation}
defines the infinite sequence of coefficients  $a_{2l}(f),~ l=1,2,...$ in the series expansion in Eq.~(\ref{1MSPnb}).

\subsection{The integral presentation and the series
	expansions of $d^2_f$ }\label{ts72}

If one uses the detailing balancing  equation~\eqref{db}, Eq.~\eqref{dm11} may be recast in the form
\begin{equation}
M_n(S)=\frac{1}{2}\int_{-\infty}^{\infty} \omega^{n}(1-e^{-\omega})Q_S(\omega)d\omega\quad n=1,3,5,...
\label{1sr}
\end{equation}
Thus, due to the fluctuation dissipation theorem~\eqref{FDT} the following sum rule is valid
\begin{equation}
M_n(S)=\frac{1}{2\pi}\int_{-\infty}^{\infty}\chi^{''}(\omega)\omega^{n}d\omega=\langle R_n(S)S\rangle_{T},\quad n=1,2,3...
\label{Mchi}
\end{equation}

Plug  the Eq.~\eqref{Mchi} in Eq.~\eqref{1MSPna}
one obtains the result
\begin{equation}
d^2_{f}(S,S)=  \left\{d^2_{BKM} +  \frac{1}{8\pi}\sum_{l=1}^{\infty}\int_{-\infty}^{\infty}\left(\frac{1}{2}\right)^{2l-1} a_{2l-1}(f)\chi^{''}(\omega)\omega^{2l-1}d\omega\right\}.
\label{redC}
\end{equation}
Formally interchanging the integration and summation in the above expression and
utilizing Eq.~\eqref{use} in the form
\begin{equation}
\sum_{l=1}^{\infty}a_{2l-1}(f) \left(\frac{\omega}{2}\right)^{2l-1}
\label{ouse}=\frac{2}{\omega}\left[g_{f}\left(\frac{\omega}{2}\right)-1\right]
\end{equation}
we get back to Eq.~\eqref{MRM1c}. Thus we have two equivalent  presentation of $d^2_f$ given by the series expansion Eq.~\eqref{1MSPna} and by the integral
presentation Eq.~\eqref{MRM1c}
.

With similar algebra, one can show that  $d^2_f$ given by Eq.~\eqref{1MSPnb} as a series expansion in terms of the even  moments of DSF $M_{p}(S)$, is equivalent to Eq.~\eqref{MRM1c}.

The alternative way to present $d^2_f(S,S)$ with the help of the even moments of DSF, Eq.~\eqref{1MSPnb}, is to start our consideration with the following (equvalent)
presentation  of Eq.~(\ref{pspMSP})
\begin{equation}
d^2_f(S,S)=\frac{1}{4}\left\{\int_{-\infty}^{\infty} \hat{g}_{f}\left(\frac{\omega}{2}\right)Q_{S}(\omega)d\omega-\langle S \rangle_T^2\right\}.
\label{MRM2}
\end{equation}
With a similar algebra as in the previous case one can show that  $d^2_f$ given by Eq.~\eqref{1MSPnb} is equivalent to Eq.~\eqref{MRM1c}.

It is important to note that the key computational advantage of both formulas Eq.~(\ref{1MSPna}) (or Eq.~(\ref{1MSPnb})), instead of Eq.~\eqref{pspMSP} (or Eq.~(\ref{spMSP})), is due to the existing presentation  of functionals $F_{n}(S;S)$ in Eq.~(\ref{VRNTa}) in terms of the iterated commutators Eqs.(\ref{R1a}). Similarly to the terminology   of Kubo for the admittance  \cite{K66} these formulas may be called  sum-rule expansions.

The presentations Eqs.~(\ref{1MSPna}) and (\ref{1MSPnb})  have an awkward feature: the summation is extended to infinity which raises the question about the convergence of the series. It is clear that the series representations should  yield a proper definition of a monotone Riemannian metric provided the corresponding convergence condition is fulfilled. This point which in general is problematic due to the unknown behavior
of the specters of the Hamiltonian $T$ and operator $S$ in functionals $F_{n}(S;S)$ (or $M_n(S)$), $n=1,2,...$ needs special examination in the framework of  concrete models.

\section[]{Application to  Models}\label{ts8}

A.)  We first illustrate the calculation of the
 whole class Riemannian
metrics on the example of the simplest but instructive system of $N$ spins in a constant  magnetic field $h$. The Hamiltonian of the model is
\begin{equation}
H(h)=\omega_0 S_z + hS_x,\qquad [S_x,S_y]=iS_z,\qquad S^2_x + S^2_y + S^2_z=S(S+1),
`	\end{equation}
where $S\geq 1/2$ is the spin quantum number. In this case the dynamical structure factor with respect to $S_x$ is
(for more details see \cite {R09} and Section 7.4.4 \cite{R13})
\begin{equation}
Q_{S_x}(\omega)=\frac{1}{2}\langle S_z\rangle_{H(0)}[\delta (\omega - \omega_0) - \delta (\omega  + \omega_0)](e^{\omega} -1)^{-1}.
\label{sh}
\end{equation}
After using the relation
\begin{equation}
\langle S_z\rangle_{H(0)}=\frac{d}{dh}\ln Z(h)=(2S+1)\coth\left((2S+1)\frac{h}{2}\right)-\coth\left(\frac{h}{2}\right)\equiv2SB_S(h/2),
\end{equation}
where $B_S(h)$ is the Brillouin function,
the application of Eq.~(\ref{sh}) to  Eq.~(\ref{MRM2}) immediately yields the result
\begin{equation}
d^2_f=SB_S(h/2)\left(\frac{\omega_0}{2}\right )^{-1}g_f\left(\frac{\omega_0}{2}\right).
\end{equation}

B.) Let us consider the Hamiltonian  \cite{LYZ10,Z13}:
\begin{equation}
{\mathcal H}(h)=k\omega\left(Q^0_k-\frac{1}{k^2}\right)+h\sqrt{k^k}(Q^+_k + Q^-_k),\quad k=1,2,...,
\label{HAQ}
\end{equation}
where  $Q^\pm_k$ are operators obeying
the commutation relations
\begin{equation}
[Q^0_k,Q^{\pm}_k]= \pm Q^{\pm}_k,\qquad [Q^+_k,Q^{-}_k]= \Phi_k(Q^0_k)-\Phi_k(Q^0_k-1),
\label{KR1}
\end{equation}
with the structure function
\begin{equation}
\Phi_k(Q^0_k)=-\Pi_{i=1}^{k}\left(Q^0_k+\frac{i}{k}-\frac{1}{k^2}\right)
\end{equation}
being a $k^{th}$-order polynomial in $k$.
The Hamiltonian (\ref{HAQ})
is employed in various physical problems (for definitions  and a partial list of references, see \cite{Z13,LYZ10}). Therefore, we shall derive closed expressions of  the monotone   Riemannian metrics within this class.

In this case our approach  is very effective since the iterative commutation
between  $T=k\omega\left(Q^0_k-\frac{1}{k^2}\right)$ and $S=\sqrt{k^k}(Q^+_k + Q^-_k)$
implies  some periodic operator structures  after a finite number of steps as it was otained in ref.~\cite{T14}:
\begin{equation}
R_n=(-1)^nR^+_{n}=\alpha^n[Q^+_k + (-1)^nQ^{-}_k], \quad (Q^-_k)^+=Q^{+}_k,
\label{sf1}
\end{equation}
indicating an analytical expression  as a function of $n$.
The parameters $k$ and $\omega $  enter in the c-number $\alpha=(k\omega)^k \sqrt{k^k}$. Thus, the  obtained series expansions Eqs. (\ref{1MSPna}) and (\ref{1MSPnb}) can be used in a rather simple  way to obtain  closed-form expressions.

The polynomial
algebra of degree $k-1$  defined  by Eqs.(\ref{KR1})
has  the following one-mode boson realization \cite{LYZ10}:
\begin{equation}
Q^+_k= \frac{1}{(\sqrt{k})^k} (b^+)^k,\quad  Q^-_k = \frac{1}{(\sqrt{k})^k} b^k.
\label{pZ}
\end{equation}
In terms of Eqs. (\ref{pZ}) the Hamiltonian of the model takes the more familiar form \cite{Z13}
\begin{equation}
{\mathcal H}(h)=\omega b^+b + h [(b^+)^k + b^k],\qquad \omega >0,\quad k=1,2,3,...
\label{HAk}
\end{equation}
where bosonic operators $b, \;b^+$   obey  the canonical commutation
relations.

The particular cases of $k = 1$ and
$k = 2$ in Eq.~(\ref{HAk}) give the Hamiltonians of the displaced and single-mode squeezed
harmonic oscillators, respectively. The Hamiltonian (\ref{HAk}) for $k=2$ is also known as Lipkin-Meshkov-Glick (LMG) model in the Holstein-Primakoff single boson representation (see e.g. \cite{Gu10} and refs. therein)
and all the result obtained here can be related to this field.

 If $p=2n$, from (\ref{Nf1}) by using the  expressions of $R_0$ and $R_{2n-1}$, we obtain
\begin{eqnarray}
F_{2n}(S;S)=-2(\omega)^{2n-1}{\mathcal K}(k),\quad n=0,1,2,...,\quad k=1,2,...,
\label{K}
\end{eqnarray}
where
\begin{eqnarray}
{\mathcal K}(k)=k^k\langle[Q^{+}_k - Q^-_k][Q^{+}_k + Q^-_k]\rangle_T,\quad
k=1,2,...
\label{MKa}
\end{eqnarray}
If $p=2n+1$, from (\ref{Nf1}) by using the  expressions of $R_0$ and $R_{2n}$, we obtain
\begin{equation}
F_{2n+1}(S;S)=2(k\omega)^{2n}{\mathcal L}(k),\quad
 n=0,1,2,...,\quad k=1,2,...,
\label{Kdr}
\end{equation}
where
\begin{equation}
{\mathcal L}(k)=k^k\langle[Q^{+}_k + Q^-_k]^2\rangle_T,\qquad k=1,2,...
\label{MKb}
\end{equation}
Evaluation of the correlation functions Eqs. (\ref{MKa}) and (\ref{MKb}) with the quadratic Hamiltonian T is now straightforward.
The results for $k=1$ and $k=2$ are:
\begin{eqnarray}
{\mathcal K}(1) &=&-1,\quad {\mathcal L}(1)=2n+1, \nonumber\\
{\mathcal K}(2) &=& -2(2n+1),\quad {\mathcal L}(2)=4n^2,
\end{eqnarray}
where $n=(e^{ \omega} - 1)^{-1}$. Taking into account the relation (\ref{VRNTa})
with the help of Eqs.(\ref{K}) and (\ref{Kdr}), the series expansions~-- Eqs.(\ref{1MSPna}) and (\ref{1MSPnb})~--  may be recast in the closed form
\begin{equation}
d^2_{f}=d^2_{BKM} + \frac{1}{4}\left(\frac{k\omega}{2}\right)^{-1}\left[1-g_{f}\left(\frac{k\omega}{2}\right)\right]{\mathcal K}(k)
\label{nu1}
\end{equation}
and
\begin{equation}
d^2_{f}=d^2_{MC} - \frac{1}{4}\left[1- \hat{g}_{f}\left(\frac{k\omega}{2}\right)\right]{\mathcal L}(k),
\label{nu2}
\end{equation}
respectively.

\section{Summary}\label{ts9}

The geometric approach traditionally plays an important role in deriving and elucidating theoretical results in various physical disciplines \cite{TF12}, especially for thermodynamics,  see e.g. Ch.~7 in \cite{R13}.
Recently,  a number of papers \cite{GY14,YH15,HHTZ16,SU16,R09,k19} advocates  an inherent relation between quantities from   information geometry  and statistical mechanics. The key point is that the monotone (contractive) Riemannian metrics (named also quantum Fisher informations) on the space of states can be identified and analyzed in terms of the behavior of the linear response functions, and vice-versa.


In the  present study,  the  quantum states we are dealing with, may be realized as a set of one-parameter family of Gibbs thermal states, Eq.~(\ref{roh}), obtained upon varying the   Hamiltonian parameter ``$h$'' conjugated to an observable ``$S$''.  Our Eq.~\eqref{MRM1c} relates the entire class of the monotone
Riemannian metrics (or Fisher informations) $d_{f}^2(S,S)$ on the set of quantum states of the system under consideration, through the (filter \cite{b}) function $g_f(x)$ to the dissipative component  of the Kubo response  function  $\chi^{''}_{S}(\omega)$ in the state $\rho$ with respect to $S$. Since, a priory  one can  choose  between many metrics,  we discuss  in Sec.V those of them which play an essential role from point of view of physics.

Practical applications of the general relations, i.e. Eqs.~\eqref{MRM1c} for the integral presentation of $d_{f}^2(S,S)$, and \eqref{redC} for the series expansion of $d_{f}^2(S,S)$ may occur in at least two aspects: to obtain estimates  by some lower and upper bounds, and to apply  the theory of perturbation.

A key point in our consideration is the use of the computational properties of the  functionals Eq.~(\ref{Nf1})  introduced previously in ref. \cite{BT11}.  These are presented  in a basis independent form as thermodynamic mean values of n-times iterated (or nested)  commutators, see Eq.~\eqref{Mchi}, between $ H(0)$ and $ \partial_h H(h)$ which in some cases provides significant  computational advantage \cite{T14}. The  functionals, Eq.~(\ref{Nf1}), are related to the moments of DSF \cite{T16} which play the role of
the well known Kubo sum rules \cite{K66}. 
Estimations of hardly computable quantities from below and above, as a tool for obtaining exact results, are also widespread and traditional in information theory and statistical mechanics. A step in
the right directions is the obtained inequalities between the different metrics obtained in Sections~\ref{ts5} and \ref{ts6}. It should be stressed  that the estimates, Eqs.~\eqref{BHFR16}, \eqref{R16}, \eqref{nNiT} and the series of inequalities in Section~\ref{ts6}, are exact and cannot be inferred from any perturbation theory.

In order to position the obtained results among  others in the field let us accent on the main source of deference.
 Note that in the definition of a metric, indeed, the essential  is the scenario that renders two states, e.g. $\rho_1$ and $\rho_2$,  distinguishable.
 One can see the difference if one compares the spectral presentation for MRM (the quantum Fisher informations) obtained through an unitary evolution $\rho(h)=e^{-ihS}\rho(0)e^{+ihS}$ (relevant to a best achievable precision in the parameter estimating problem) with the spectral presentation  obtained  by infinitesimally variation of a control parameter ``$h$''  in the Hamiltonian  that parameterized the  Gibbs thermal states (relevant to quantum criticality). More technically,  differences appear caused by  the additional term  $|E_n-E_m|^2$ (using our notations) in the denominator of the corresponding summand of the Bures metrics. This causes differences in the corresponding  relations
  with the response functions (c.f. with ref.~\cite{SU16}).

  Assuredly, links between families of metrics over the manifold of density matrices  and response functions, obtained from  different viewpoints,  might cross-fertilize and extend both directions.

 \section*{Acknowledgements}

 I would like to thank Valentin Zagrebnov for useful discussions  and valuable comments.
 This work was supported by a Grant with the Joint
 Institute for Nuclear Research, Dubna, Russian Federation -- THEME
 01-3-1137-2019/2023 and Grant No~D01-378/18.12.2020
 of the Ministry of Education and Science of Bulgaria.


\def\thesection{Appendix}
\appendix


\section{\mbox{}\hspace*{-4mm}: Operator Monotone Functions} \label{tAA}

Let us recall the definition of the operator monotone function. A real value function $f: (a,b)\to \mathbb{R}$, where $(a,b)\in\mathbb{R} $, is said to be  monotone for finite-size $n\times n$ matrices $A \leq B$, whenever  $A$ and $B$ are self-adjoint and their  eigenvalues are in $(a,b)$, if $f(A) \leq f(B)$. If a function is monotone for every matrix size, then it is called {\it operator monotone}. 
By approximation arguments this definition is extend-able for operators on an infinite dimensional Hilbert space (for a more information about operator monotone functions, see e.g. \cite{HP14,Ch15}).

Examples of  the operator monotone functions $f\in \mathcal{F}_{op}$ are given in the list \cite{P96,HP14,F08}:
\begin{equation}
\begin{split}
&f_{Har}(x)=\frac{2x}{x+1},\quad f_B(x) = \frac{x+1}{2},\quad f_{BKM}(x)=\frac{x-1}{\ln x}, \\
& f_{MC}(x)=\left(\frac{x-1}{\ln x}\right)^2\frac{2}{1+x},\quad f_{G}(x)=\sqrt{x}, \\
& f_{WYD}(\alpha, x) = \alpha(\alpha-1)\frac{(x-1)^2}{(x^{\alpha}-1)(x^{1-\alpha}-1)},\quad 0<\alpha<1.
\end{split}
\label{uga}
\end{equation}
In addition, we shall state  a class of operator monotone functions  which are an
 useful tool to illustrate ideas of this paper. The function $f_p(x)$
\begin{equation}
f_p(x)=\frac{p-1}{p}\left(\frac{x^p-1}{x^{p-1}-1}\right)
\label{clmf}
\end{equation}
is an operator monotone function for $-1\le p\le 2$. Note that  a part of the above given examples of operator monotone functions, Eq.~\eqref{uga}, belongs to this class (see below) and in addition one can show that
\begin{equation}
f_G(x)\le f_p(x)\le f_{B},\quad \frac{1}{2}\le p \le 2.
\end{equation}

An operator monotone function $ f:\mathbb{R}^+ \to \mathbb{R}^+$ is called symmetric if  $f(x):(0,+\infty) \rightarrow (0,+\infty); f(x^{-1})=f(x)/x$  and normalized  if $f(1)=1$. It is commonly accepted to denote by $\mathcal{F}_{op}$ the set of the functions $f(x)$ characterized in this way. Some time these functions are called standard operator monotone functions as well \cite{HP14}.
The above examples of operator monotone functions have exactly these properties.
The function $f_{Har}(x)$ is the minimal, while $f_{B}(x)$ is the maximal operator monotone functions on $[0,+\infty)$. The former  defines a metric  known as RLD metric while the last one gives rise to the SLD   metric (named also Bures metric  or fidelity susceptibility). The   function $f_{BKM}(x)$  leads to the  Bogoliubov-Kubo-Mori  metric.
The function $f_{WYD}(x)$ is associated to the Wigner-Yanase-Dyson metric.  The function $f_{MC}(x)$
was first conjectured in the paper \cite{MC90}, which explains here the subscript $MC$.  Its matrix monotonicity was proved in \cite{P96}.

In according with the Kubo and Ando theory of operator means (see, \cite{KA80}, Chapter V in ref. \cite{HP14}, \cite{Ch15}) to every operator mean corresponds a unique operator monotone function $f\in \mathcal{F}_{op}$, and conversely. If $f\in \mathcal{F}_{op}$ then the corresponding mean is given by Eq.~(\ref{KA}).
The function Eq.~\eqref{clmf}  is related with the so called power difference means, where the values $p=-1,1/2,1,2$ correspond to the mentioned above operator monotone functions:
$  f_{-1}(x)\equiv f_{Har}(x)$ to harmonic mean, $f_{1/2}(x)\equiv f_{G}(x)$ to geometric mean, $
f_1(x)=f_{BKM}(x)$ to logarithmic mean, and $f_2(x)=f_{B}(x)$ to arithmetic mean, respectively.

\end{document}